\documentclass[12pt]{article}

\usepackage[utf8]{inputenc}
\usepackage{graphicx}
\usepackage{amsmath, amssymb}
\usepackage{geometry}
\usepackage{authblk}
\usepackage{cite}
\usepackage[colorlinks=true, linkcolor=blue, citecolor=blue, urlcolor=blue]{hyperref}
\usepackage{setspace}
\usepackage{titlesec}
\usepackage{abstract}

\titleformat{\section}{\normalfont\large\bfseries}{\thesection.}{1em}{}
\titleformat{\subsection}{\normalfont\normalsize\bfseries}{\thesubsection.}{1em}{}

\title{\textbf{\Large Uncovering the network geometry of green bonds}}

\author[1]{Xinyue Zhang}
\author[1]{Alexander P. Kartun-Giles}

\affil[1]{Institute for Sustainable Resources, Bartlett School of Environment, Energy and Resources,\\
University College London, WC1H~0NN, UK}

\date{}

\begin{document}

\maketitle

\begin{abstract}
With the rapid growth of the green bond market amid increasing emphasis on sustainable development, understanding its structural properties and potential systemic risks has become essential. This study applies Minimum Spanning Tree (MST) and Hierarchical Tree (HT) methods introduced by Mantegna, which gives a minimum spanning tree of bonds whose mutual distance corresponds to the correlation coefficient of the synchronous time evolution of the difference of the logarithm of their monthly price. Our analysis reveals that green bonds exhibit tighter ultrametric distances on average compared to stocks and conventional bonds, indicating a highly interconnected market structure. We also identify strong sectorial clustering, with green bonds in the utilities sector emerging as the most central nodes by average total betweenness centrality, suggesting their critical role in potential risk propagation. These findings highlight both the opportunities and vulnerabilities inherent in the green bond market, offering insights for investors and policymakers on monitoring concentration, enhancing transparency, and mitigating systemic risks as the market continues to evolve.
\end{abstract}

\section{Introduction}

With the continuous increase in global attention to environmental issues and the strong support of national policies for green industries, the sustainable development strategy has generated a huge demand for environmentally friendly investments. There is still a considerable gap between the current fundraising efforts to effectively address climate change and the actual funds required \cite{pham2016,buchner2023}. Against this backdrop, green bonds have emerged as an innovative financial instrument in 2007, aiming to open up a dedicated financing channel for sustainable and environmentally friendly projects and thus play a crucial role in promoting the global economic transformation towards green \cite{flammer2021}. 

However, the relevant laws and market scale of green bonds are still in the developmental stage. Currently, green bonds face issues such as insufficient contract protection for investors, problems regarding transparency and reporting metrics, pricing issues, and the concern of greenwashing \cite{doran2019}. Overall, green bonds prices become correlated, not just due to interest rates, but sectorially and nationally as investor sentiment changes. Network geometry, in particular the method introduced by Mantegna in 1999 \cite{mantegna1999hierarchical} provides a framework for understanding the correlation of prices in financial markets and how the bonds can form an interesting geometric structure. 

The remainder of the paper is structured as follows. In Section~\ref{sec:2} we present a literature review looking at Mantegna’s paper for Minimum Spanning Tree (MST) and Hierarchical Tree (HT) methods, looking at gaps in past studies to shape the analytical framework. In Section~\ref{sec:3} we discuss our methodology for MST and HT construction and visualizations, engineered to dissect the market's network structure, revealing bond and participant interconnections. In Section~\ref{sec:4} we present our results concerning the empirical findings on the network and hierarchical structure of the green bond market, and in Section~\ref{sec:5}, we conclude.

\begin{figure}[t!]
    \centering
    \includegraphics[width=1\textwidth]{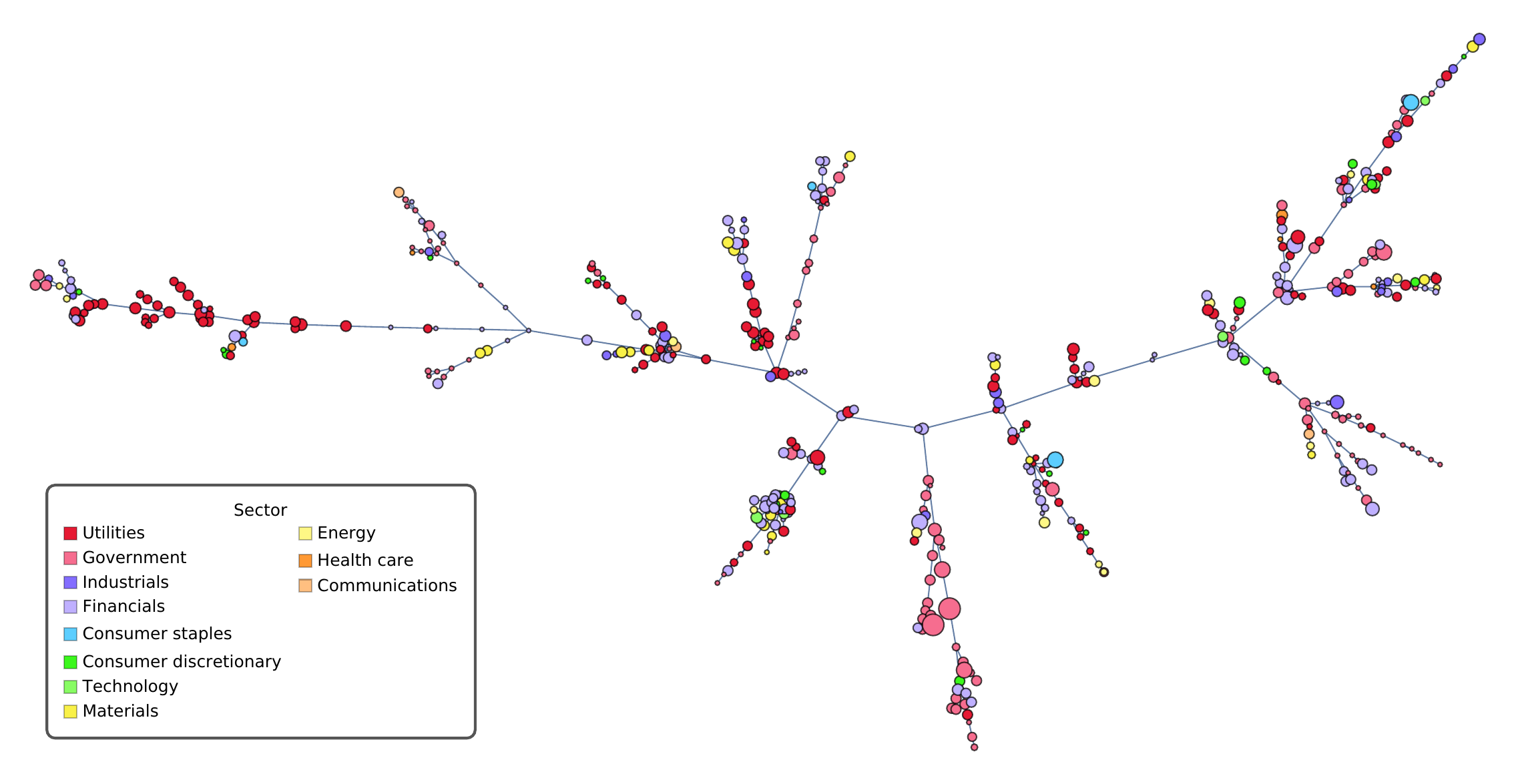}
    \caption{The minimum spanning tree based on price correlations of all green bonds in the dataset. Node size indicates the issuance size in USD, scaling linearly, and the node colours represent the sectors in which the bonds are issued.}
    \label{fig:1}
\end{figure}

\begin{figure}[t!]
    \centering
    \includegraphics[width=1\textwidth]{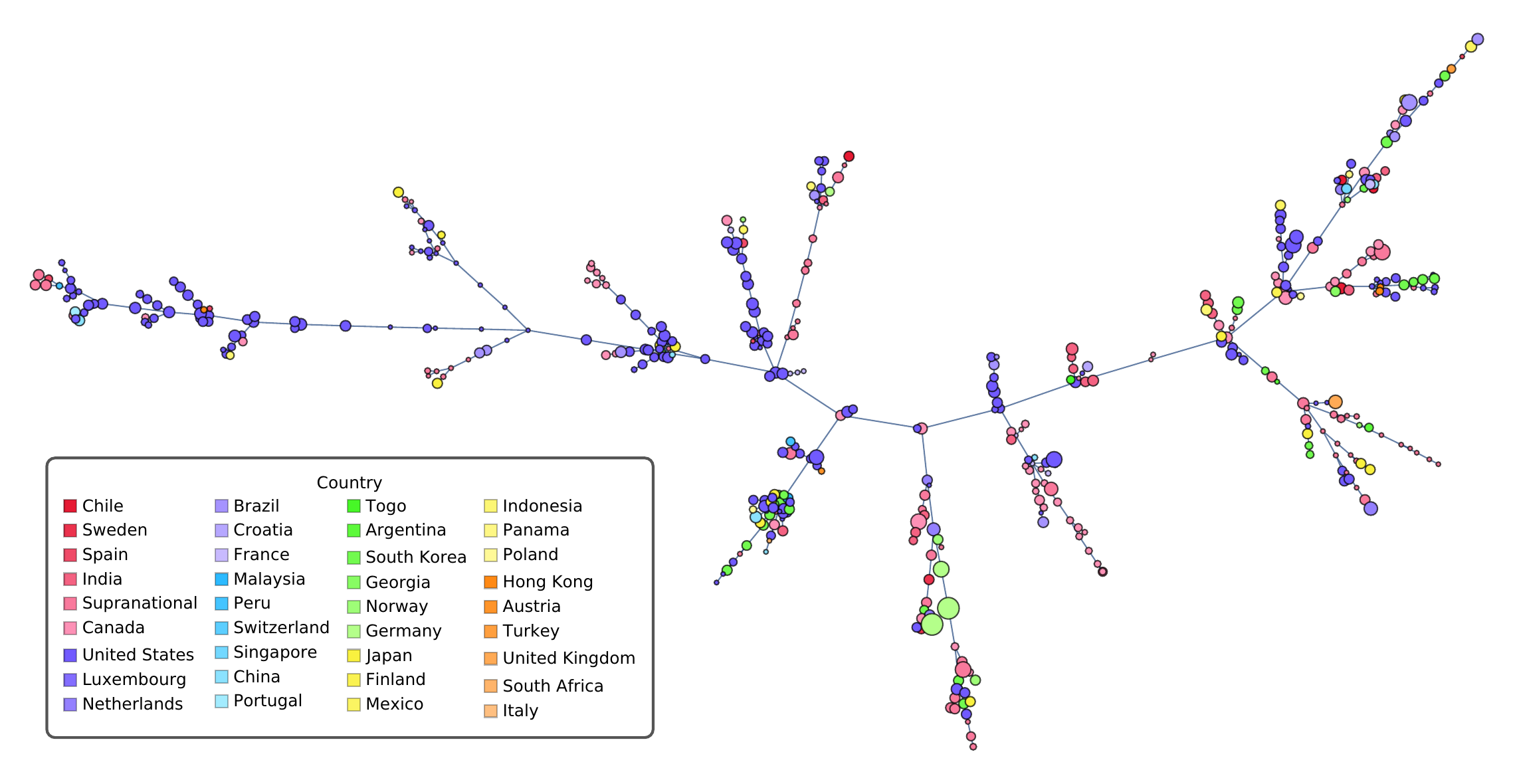}
    \caption{The minimum spanning tree based on price correlations of all green bonds in the dataset. Node size indicates the issuance size in USD, scaling linearly, and the node colours represent the issuance country.}
    \label{fig:2}
\end{figure}

\section{Literature Review}\label{sec:2}

Mantegna’s \cite{mantegna1999hierarchical} study, \textit{Hierarchical Structure in Financial Markets} has been a cornerstone in the analysis of financial market networks. Mantegna applied the MST and HT methods to dissect stock portfolios from the S\&P 500 and Dow Jones Industrial Average spanning from July 1989 to October 1995. He found that stocks within specific sectors such as capital goods, retail, and food and tobacco had a tendency to cluster by industry. This clustering indicated the existence of inter-linkages among diverse stocks. Mantegna's objective was to establish a coherent economic taxonomy for stocks traded in financial markets. Through this, he managed to disclose how stocks in the US markets inherently clustered based on shared attributes such as industry, project location, or other characteristics. Consequently, this assisted in uncovering the common economic factors that impacted their price fluctuations. The role of the MST method in this study was of paramount importance. It functioned to streamline complex network architectures. Within these simplified structures, specific nodes emerged as essential connectors, which were crucial for comprehending the relationships and dynamics within the financial market network.

Understanding links between network science and sustainable investment has seen significant recent attention \cite{reboredo2020network, karim2022dependence, ferrer2024interdependence, chen2024timefrequency, wang2024firm, kartun-giles2023introduction}. There is still a gap in the research using MST and HT to study the green bond market. Halkos, Managi, and Tsilika \cite{halkos2020} constructed a relational network between 53 organizations and 96 national market participants, spotlighting international flows and interdependencies among countries and regions. However, this approach has several limitations. Merely aggregating the weights of network edges might disregard the intricacy of multilayered relationships within the network, resulting in an inadequate comprehension of the interdependence between countries.

Although centrality metrics offer a means to gauge a country’s standing within the network, they might not comprehensively capture the heterogeneous roles of certain countries in the green bond market, especially those that contribute substantially through indirect channels or secondary markets. Moreover, the overly convoluted methodology and sequencing have diminished the readability of the study. In contrast to the methods utilized by Halkos, Managi, and Tsilika \cite{halkos2020}, MST proffers benefits over traditional financial approaches by furnishing a simplified portrayal of all potential interconnections within a network. This simplification empowers professionals to more lucidly apprehend the system’s connections and pinpoint core nodes, thereby augmenting the readability and comprehensibility of the internal network.

\section{Data and Methodology}\label{sec:3}

\begin{figure}[t!]
    \centering
    \includegraphics[width=1\textwidth]{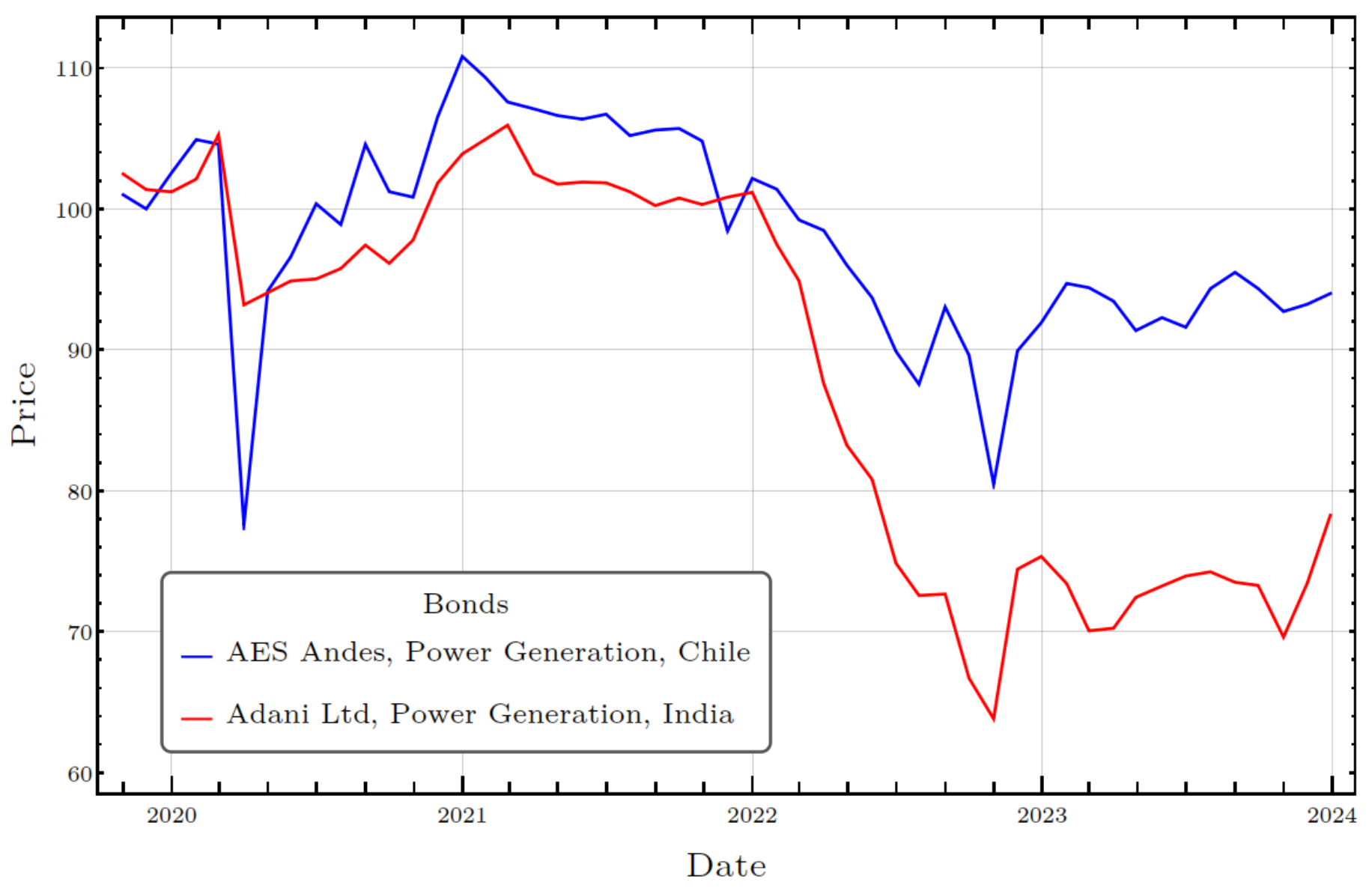}  
    \caption{Two green bond price time series that show strong correlation, and overlap in time significantly. Once we have these time series for every pair of bonds, we can measure their correlation and build the corresponding minimum spanning tree as discussed in Section~\ref{sec:2}.}
    \label{fig:3}
\end{figure}
\begin{figure}[t!]
    \centering
 \includegraphics[width=1\textwidth]{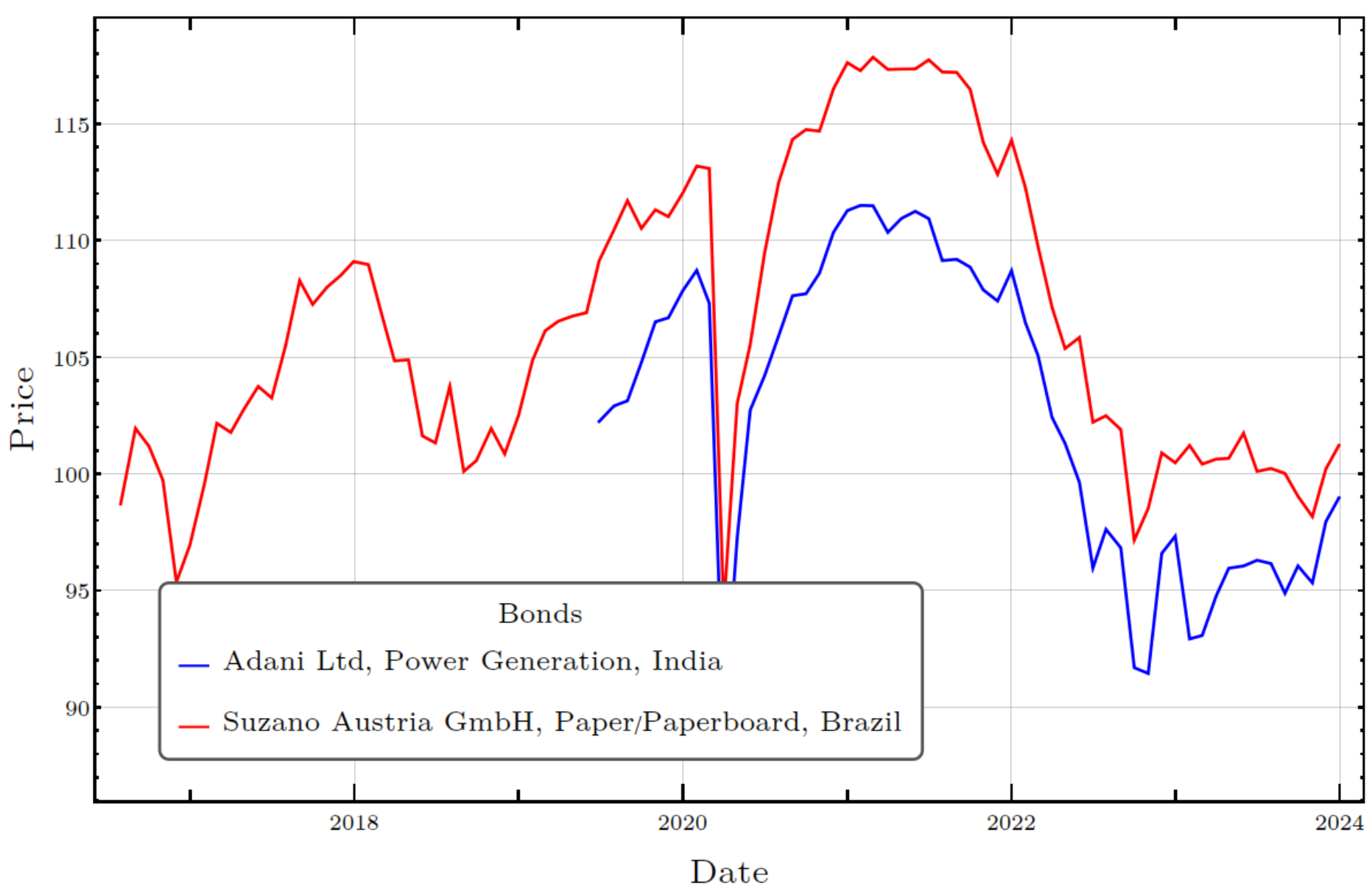}
    \caption{Two green bond price time series do not always overlap. Once overlap is less than 5 dates, we remove this link from the complete graph on the bonds, and then using the corresponding metric, build the MST.}
    \label{fig:4}
\end{figure}

We obtained a unique dataset from the London Stock Exchange Group (LSEG), comprising Bond ID, time, price index, and other relevant information. The initial dataset encompassed 572 global green bonds from March 2011 to December 2023. After excluding 94 bonds with missing price index data and further filtering for bonds with at least two years of data, 478 bonds were retained. We deliberately selected two crucial periods: 2020--2021 (with 32 bonds) and 2022--2023 (with 203 bonds). The former represents the nascent stage when green bonds were emerging, while the latter corresponds to a more established and widespread phase. This selection enables a comparison of the market's evolution.

The time series for two bonds do not overlap as the bonds can be issued or mature at different dates. To avoid observing correlation coefficients based on limited mutual data, we remove the links for the weighted complete graph on which the MST is built when the corresponding time series overlap at strictly less than 5 dates.

Minimum Spanning Tree (MST) and Hierarchical Tree (HT) methodologies, originally proposed by Mantegna \cite{mantegna1999hierarchical} in the field of econophysics, are employed to investigate the green bond market. The network method is a fundamental approach that uses simple correlations as a starting point \cite{yagli2023complex}. The MST is then built using the correlation coefficient as the basis of metric, and we can then observe the clustering within sectors and nationalities, and to observe the typical distance between the bonds. We show two time series in Figures~\ref{fig:3} and~\ref{fig:4}. The use of a metric on bonds links these correlations to network \textit{geometry}, which has seen wide ranging research in complexity science and engineering, see e.g. \cite{boguna2021}, 

The correlation coefficient is a statistical measure used to assess the strength and direction of the linear relationship between two variables. It varies between $+1$ (indicating complete positive correlation) and $-1$ (complete negative correlation). A value of 0 denotes no linear correlation between two bonds.

Let $\vec{r}_i(t)$ represent the time series of log-returns or yield rates for green bond $i$ at time $t$. The correlation coefficient $\rho_{ij}$ between green bonds $i$ and $j$ is computed using:

\begin{equation}
\rho_{ij} = \frac{\langle \vec{r}_i \vec{r}_j \rangle - \langle \vec{r}_i \rangle \langle \vec{r}_j \rangle}
{\sqrt{\left( \langle \vec{r}_i^2 \rangle - \langle \vec{r}_i \rangle^2 \right) \left( \langle \vec{r}_j^2 \rangle - \langle \vec{r}_j \rangle^2 \right)}}
\end{equation}

where $\langle \cdot \rangle$ denotes the time average over the observation window. Let $P_i(t)$ denote the closing price of green bond $i$ at time $t$, and define log returns as:

\begin{equation}
r_i(t) = \ln P_i(t) - \ln P_i(t - 1)
\end{equation}

Using log-returns helps eliminate scale effects and makes bonds with different price levels more comparable. However, because correlation coefficients do not satisfy the properties of a metric (such as the triangle inequality), they cannot be used directly as distance measures. Therefore, we transform them into distances using:

\begin{equation}
d_{ij} = \sqrt{2(1 - \rho_{ij})}
\end{equation}

Once the pairwise distances $d_{ij}$ between green bonds are obtained, Kruskal’s algorithm is used to construct the MST connecting the $N$ green bonds in the portfolio \cite{brida2010use}. This involves iteratively connecting the pair of bonds with the smallest distance while ensuring no cycles are formed. The procedure continues until all nodes are connected, resulting in a tree of $N-1$ edges.

To visualize the MST, we applied the Kamada–Kawai layout algorithm \cite{kamada1989algorithm}, which minimizes spring forces and produces a visually interpretable graph that reflects the relational geometry of the bonds. Hierarchical Tree (HT) analysis is built upon the subdominant ultrametric distance. For two nodes $i$ and $j$, the subdominant ultrametric distance $d^{\mathrm{ult}}_{ij}$ is defined as the maximum Euclidean distance along the shortest path between $i$ and $j$ in the MST. Both correlation-based classifications (via HT) and geometric topology (via MST) are obtained from the tree structures. The HT method reveals nested clusters in the network and allows a detailed classification of correlations between green bonds, shedding light on the hierarchical structure of the market, the clustering and centrality of issuers, and possible economic taxonomies.

\section{Results}\label{sec:4}

Figures~\ref{fig:1} and~\ref{fig:2} shows the network is becoming more consolidated and interconnected over time, with nodes clustering more densely. Specifically, from 2020 to 2021, only 38\% of pairwise distances between nodes were less than 0.3. However, by the end of 2023, this figure had risen to 66\%. Mantegna’s paper observed that distances between stocks typically exceeded 0.95, underscoring the unique connectivity of the green bond market's current state. A short correlation distance between bonds reflects synchronized price movements across the market - a potentially hazardous situation. In a tightly connected network, the collapse of a central node due to greenwashing or other factors can cause systemic risk, and the lack of contract clauses to protect investor interests could severely damage investor confidence. This price collapse will extend to various nodes across the network.

\begin{figure}[h!]
    \centering
    \includegraphics[width=1\textwidth]{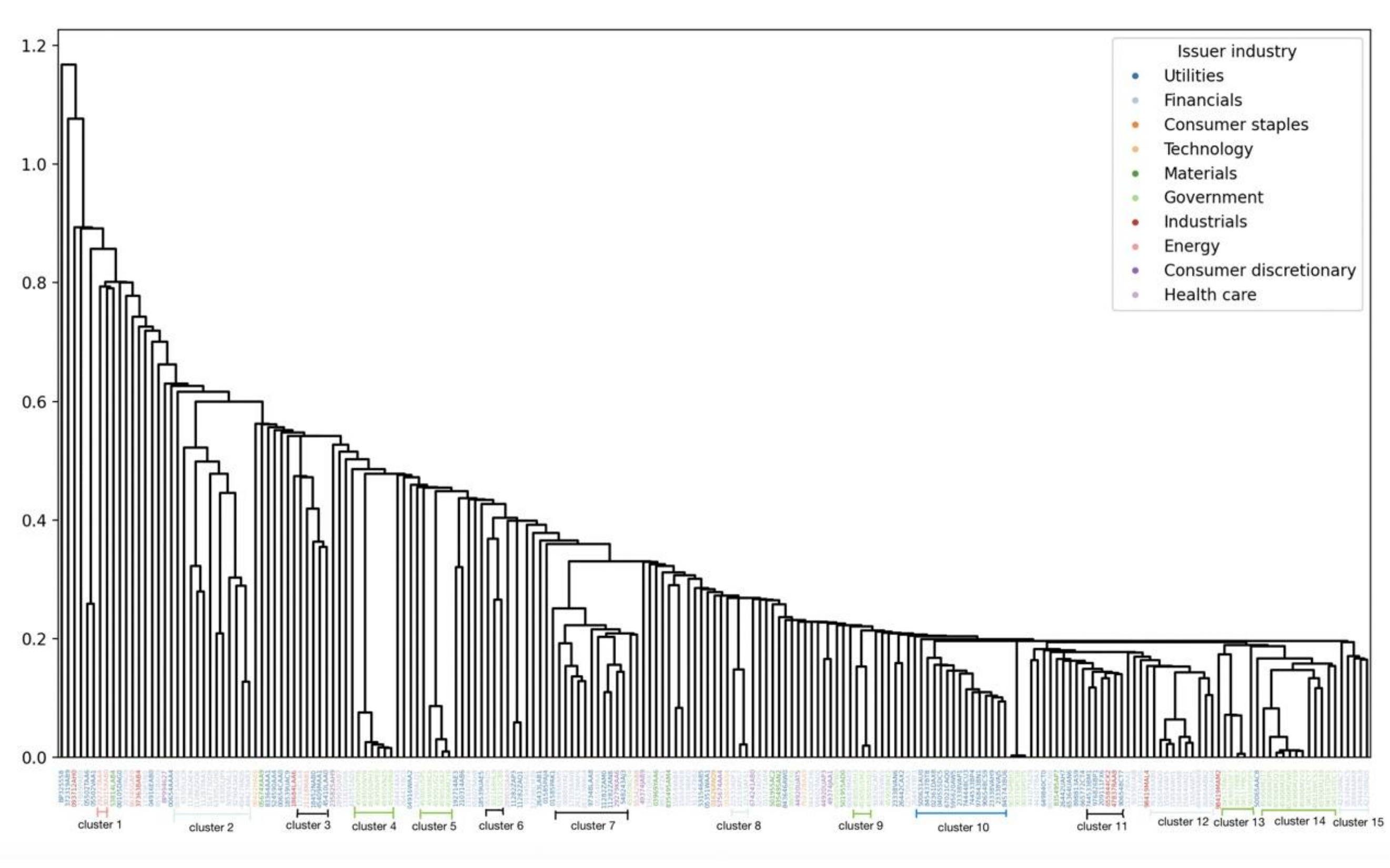}
    \caption{HT of 203 green bonds (2022--2023). Bond IDs are color-coded by issuer industry. Numbered clusters highlight regions of high similarity. The structure reflects industry concentration and correlation strength. The list of bonds used can be found in the supplementary material.}
    \label{fig:5}
\end{figure}

The green bond market is notably concentrated in three issuing sectors: utilities, government, and financials (see Figs.~\ref{fig:4} and~\ref{fig:5}). The betweenness centrality of the bonds in these sectors exceeds the others by an order of magnitude, with Utilities the overall leader, see Table~\ref{tab:betweenness}. When we take into account the number of bonds in each sector, considering instead the mean centrality of a bond in each sector, the utilities sector takes the lead, see Table~\ref{tab:t2} suggesting that despite the large issuance and popularity of green government bonds, its is the utilities sector which leads the hierarchical structure of green bond prices. This may be due to the liquidity of government bonds, which dampens their influence in affecting the price of other securities. These projects frequently involve renewable energy (e.g., solar, wind) and energy efficiency, creating aligned investor expectations and responses, resulting in higher price correlations.

The financial sector (these are bonds issued by banks to various projects), shows a similar pattern. Many green bond proceeds are allocated to renewable energy, green buildings, and energy efficiency—currently key pillars of sustainable investment. The similarity in use of proceeds across issuers enhances network connectivity. This sectoral homogeneity is a major contributor to the high correlation levels observed in the green bond market, driving the tightly connected structure in MSTs and HTs.


Government green bonds have become increasingly central in the green bond network, driven by climate goals and the breadth of projects they fund. For example, Canada's updated Green Bond Framework (2023) supports clean transport, energy (including nuclear), biodiversity, and natural resource management. Government-issued bonds are attractive due to their strong credit ratings, transparency, and reliability. This investor preference reinforces their network centrality. Since 2021, sovereign green bonds have grown rapidly, representing 24\% of all issuances. Canada’s inaugural green bond in 2022, with final orders reaching CAD 11 billion, exemplifies this trend.

As seen in Fig.~\ref{fig:5}, clusters 4, 5, and 13 consist almost entirely of bonds from the same issuer. These bonds share characteristics like credit ratings, maturity, and coupon structures, and hence exhibit highly correlated price movements. This high volume issuance enhances liquidity and further strengthens government bonds' prominence within the green bond network.

We summarise our results as follows.

\begin{enumerate}
    \item \textbf{High market connectivity:} Green bonds exhibit significantly shorter ultrametric distances than the stocks observed in Mantegna's study, suggesting strong market connectivity.
    
    \item \textbf{Sectorial clustering:} Bonds issued in the \emph{Government}, \emph{Utilities}, and \emph{Financials} sectors dominate the network, showing the highest clustering and betweenness centrality values, often exceeding other sectors by an order of magnitude.
    
    \item \textbf{National issuance patterns:} When distinguishing bonds by issuing country, \emph{U.S. green bonds} consistently occupy the most central positions by betweenness centrality in the network, though they are also the most common issuance nation in our dataset given the need for available price time series.
    
    \item \textbf{Country-level clustering:} Bonds from the same issuing nation frequently form tight clusters, reflecting similarities in credit ratings, maturity structures, and policy frameworks that drive synchronized price movements.
\end{enumerate}

\begin{table}[t!]
\centering
\caption{The total betweenness centrality of all bonds by sector.}
\begin{tabular}{l r}
\hline
\textbf{Sector} & \textbf{Betweenness Centrality} \\
\hline
Consumer staples        & 0 \\
Communications          & 912 \\
Technology              & 3,163 \\
Energy                  & 4,561 \\
Industrials             & 5,919 \\
Health care             & 9,860 \\
Consumer discretionary  & 27,317 \\
Materials               & 28,910 \\
Government              & 348,846 \\
Utilities               & 400,770 \\
Financials              & 444,875 \\
\hline
\end{tabular}
\label{tab:betweenness}
\end{table}

\begin{table}[h!]
\centering
\caption{Average betweenness centrality for bonds by sector.}
\begin{tabular}{l r}
\hline
\textbf{Sector} & \textbf{Average Betweenness Centrality} \\
\hline
Consumer staples        & 0.000 \\
Energy                  & 147.129 \\
Communications          & 228.000 \\
Industrials             & 257.348 \\
Technology              & 632.600 \\
Consumer discretionary  & 827.788 \\
Materials               & 1204.580 \\
Health care              & 1643.330 \\
Government               & 2603.330 \\
Financials               & 2616.910 \\
Utilities                & 2946.840 \\
\hline
\end{tabular}
\label{tab:t2}
\end{table}

\section{Conclusion}\label{sec:5}

This paper has examined the network geometry of the green bond market using Minimum Spanning Tree (MST) and Hierarchical Tree (HT) methods, enabling a detailed exploration of correlations, clustering, and systemic structures within the market. Our results show that green bonds exhibit much shorter ultrametric distances than those observed in stock networks \cite{mantegna1999hierarchical}, implying a tightly connected market with strong synchronization of price dynamics. We also find clear evidence of sectorial clustering, with Government, Utilities, and Financial bonds dominating both the total and average betweenness centrality. These sectors, alongside U.S.-issued bonds, emerge as the key structural pillars of the green bond network, reflecting their central role in price transmission and potential risk propagation.

Understanding these features is essential for both market participants and policymakers. For investors, knowledge of clustering patterns and central nodes provides insight into diversification limits and the potential for systemic contagion in the event of market stress or greenwashing scandals. For regulators and governments, identifying the sectors and issuers that form the backbone of the green bond network can inform the design of transparency requirements, disclosure standards, and market-stabilizing interventions. As the market expands, maintaining its integrity and resilience will be vital to ensuring that green bonds fulfill their role as a cornerstone of sustainable finance. Beyond this article, we can begin to investigate this in greater detail on more complete datasets, with the potential for a more rigourous classification of the network geometry in this case, with greater comparison with vanilla bonds and other markets.

Given these early results, future research could extend this analysis by observing the impact interest rates can have on green bonds which may present stronger correlations than are actually present between two securities. By observing non-trivial (i.e. unexpected) correlations between these products once we have filtered off the background forces that can correlated these products, there is genuine promise to observe the inter-connections between green bonds internationally in a way that can advise macro-prudential regulation, and protect the market from an undetected systemic crises.

\section*{Acknowledgements}
We thank Nadia Ameli, Denitsa Angelova, Fabio Caccioli, Carl Dettmann, Max Falkenberg, Maurizio Fiaschetti, Michael Grubb, Sumit Kothari, Andrea Macrina, and Jamie Rickman for many helpful discussions. We acknowledge support from the European Research Council (ERC) under the European Union’s Horizon 2020 research and innovation programme (grant agreement No 802891).


\end{document}